\documentclass[clsnofoot,10pt,twocolumn,letterpaper]{IEEEtran}
\usepackage{graphicx}
\usepackage{amsmath}
\usepackage{amssymb}
\usepackage{algorithm}
\usepackage{algorithmic}
\usepackage{array}
\usepackage{multicol}
\usepackage{multirow}
\usepackage{soul}

\begin{document}
\title{User Selection in Millimeter Wave Massive MIMO System using Convolutional Neural Networks}
\author{Salman Khalid, Waqas~bin~Abbas, Farhan Khalid, \textit{Member, IEEE}, Michele Zorzi, \textit{Fellow, IEEE}
	\thanks{
		Salman Khalid (corresponding author, email: salmankhalid16@yahoo.com), W.~bin~Abbas and Farhan Khalid are with the National University of Computer and Emerging Sciences (NUCES), Islamabad, Pakistan. M. Zorzi is with the Department of Information Engineering, University of Padova, Italy.}
	}
\maketitle
\begin{abstract}
A hybrid architecture for millimeter wave (mmW) massive MIMO systems is considered practically implementable due to low power consumption and high energy efficiency. However, due to the limited number of RF chains, user selection becomes necessary for such architecture. Traditional user selection algorithms suffer from high computational complexity and, therefore, may not be scalable in 5G and beyond wireless mobile communications. To address this issue, in this letter we propose a low complexity CNN framework for user selection. The proposed CNN accepts as input the channel matrix and gives as output the selected users. Simulation results show that the proposed CNN performs close to optimal exhaustive search in terms of achievable rate, with negligible computational complexity. In addition, CNN based user selection outperforms the evolutionary algorithm and the greedy algorithm in terms of both achievable rate and computational complexity. Finally, simulation results also show that the proposed CNN based user selection scheme is robust to channel imperfections.    
\end{abstract}
\section{Introduction}
Partially connected hybrid structures for millimeter wave (mmW) massive MIMO systems are drawing considerable attention from researchers due to the benefits of spatial multiplexing coupled with low complexity, low power consumption and high energy efficiency \cite{ref_01}, \cite{ref_02}. However, such systems can only accommodate a limited number of users due to restrictions on the number of RF chains \cite{ref_03}. In the environments where the number of users is large compared to the number of RF chains, user selection is essential. In the existing literature, many model driven approaches are proposed for user or antenna selection. The performance of optimization and model driven methods for user selection is evaluated in terms of either bit-error-rate (BER) or achievable data rate. Evolutionary algorithms for user selection are employed in \cite{ref_04} to find a suboptimal solution with a reduced computational complexity compared to user selection based on exhaustive search. A greedy algorithm for user selection with downlink beamforming is applied in \cite{ref_05} to further reduce the computational complexity, but at the cost of a reduction in achievable rate. User selection based on matching theory is proposed in \cite{ref_06}. A method based on an Adaptive Markov Chain Monte Carlo (AMCMC) algorithm for multi-cell multi-user massive MIMO down-link systems is proposed in \cite{ref_07}. The authors in \cite{ref_08} utilized a generalized power iteration precoding (GPIP) algorithm to generate a joint solution for user selection, power allocation, and downlink precoding. However, all these algorithms are sub optimal in nature. The optimal solution to the user selection problem can be obtained by employing an exhaustive search (ES) algorithm which enumerates all possible combinations of active users, but leads to an exponential growth of the complexity with the increase of the number of active users. Thus, the ES algorithm is not suitable for massive MIMO systems due to high computational complexity. 

Recently, data driven deep learning (DL) methods have shown great potential in dealing with the problems of channel estimation \cite{ref_09}, \cite{ref_10} and RF precoder design \cite{ref_11}. Similarly, data driven methods were applied to the transmit antenna selection problem, to obtain the best antenna subset \cite{ref_11}, \cite{ref_12}. Based on the availability of appropriate datasets, classifiers such as k-nearest neighbors (KNN) and support vector machine (SVM) can be trained to map the available channel dataset into the selected antenna subset. The author in \cite{ref_12} has proved that the performance of KNN and SVM based classifiers exceeds that of conventional optimization driven methods. Compared to optimization driven methods, data driven models are computationally efficient as well, which makes them convenient to be implemented in real-life applications. Note that, even though data driven methods have been widely applied to address multiple problems \cite{ref_09}-\cite{ref_12}, no work has been done so far on their application to user selection. Therefore, inspired by the efficiency and great potential of data driven techniques, in this letter we apply DL based convolutional neural networks (CNN) to the user selection problem in a massive MIMO wireless communication system. Our motivation can be explained as follows:
\begin{itemize}
	\item For the user selection problem, it is well known that the computational complexity of the ES algorithm grows exponentially with the number of active users. Hence, this algorithm is not appropriate for implementation in a real time environment.
	\item Traditional user selection algorithms require many serial iterations which generally result in a high complexity. However, in a neural network (NN), the training phase is generally performed offline, and therefore the online deployment has very low complexity and is limited to a few matrix multiplications and additions.
	\item The availability of a large dataset and the use of a huge number of iterations in the training phase enable CNNs to understand the complicated features of wireless channels.
\end{itemize}
\subsection{Novelty and Contribution}
In this letter, we propose a DL-based user selection design approach and develop a CNN which is trained using the channel realizations as the dataset and learns how to optimize the user selection to maximize the sum rate. The contributions can be summarized as follows:
\begin{itemize}
	\item \textit{New Design Approach}: A CNN based technique is proposed to directly perform the user selection without the involvement of any iterative algorithm after the CNN has been trained. The dataset is obtained after performing exhaustive search on each channel realization to select a subset of the active users which maximizes the sum rate. It is pertinent to mention that both training and dataset generation are performed offline.
	\item \textit{Robustness to imperfect CSI}: The proposed algorithm ensures robustness to imperfect channel state information (CSI). Firstly, during offline training, the CNN learns how to optimize the user selection to approach the ideal sum rate using practical channel estimates. Secondly, during the online deployment, the CNN can itself adapt to imperfect CSI and achieve robust performance in the presence of channel estimation errors.
\end{itemize}
\section{System Model}
\begin{figure}[tbph]
\centering
\includegraphics[height=1.8in,width=3.4in]{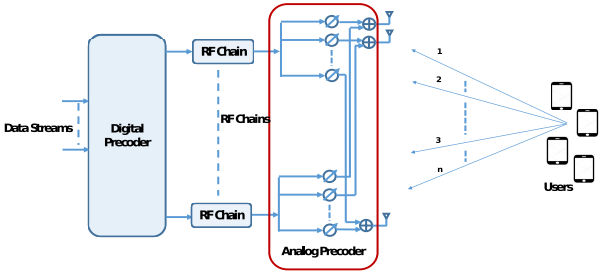}
\caption{System Model for User Selection}
\label{fig:sysmod}
\end{figure}
We consider a massive MU-MIMO downlink system with a transmitter equipped with $N_{T}$ transmit antennas, $N_{s}$ data streams and $N_{RF}$ RF chains serving $N_{R}$ single antenna users. We assume a multi-user precoding case where the transmitter communicates with each user via only one stream. Hence, the number of data streams is equal to the number of users. Further, we assume that the maximum number of users that can be simultaneously served by the transmitter equals the number of RF chains. We assume that $N_{R}$ $>$ $N_{RF}$. Hence, we have to choose a subset of $N_{r}$ users so that $N_{RF}$ = $N_{r}$. Our proposed CNN method will select $N_{r}$ users out of the available $N_{R}$ users and afterwards we will perform hybrid precoding for the selected users. For the sake of simplicity, let $N_{s}$ = $N_{RF}$ = $N_{r}$ = $N$. On the downlink, the transmitter applies an $N \times N$ baseband precoder $\mathbf{F}_{BB} = [\mathbf{f}^{BB}_{1}, \mathbf{f}^{BB}_{2}, ..., \mathbf{f}^{BB}_{N}]$ followed by an $N_{T} \times N$ RF precoder $\mathbf{F}_{RF} = [\mathbf{f}^{RF}_{1}, \mathbf{f}^{RF}_{2}, ..., \mathbf{f}^{RF}_{N}]$. The transmitted signal is represented as
\begin{equation}
    \mathbf{x} = \mathbf{F}_{RF} \mathbf{F}_{BB} \mathbf{s}
\end{equation}

where $\mathbf{s} = [s_{1}, s_{2}, ..., s_{N}]^{T}$ is the $N \times 1$ transmitted signal vector. Furthermore, $\mathbf{F}_{RF}$ and $\mathbf{F}_{BB}$ are evaluated using the algorithm proposed in \cite{ref_03}. We have adopted a narrowband block fading channel \cite{ref_13}, where the \textit{n}th user observes the received signal as
\begin{equation}
    \mathbf{r}_{n} = \mathbf{h}_{n} \sum_{u = 1}^{N} \mathbf{F}_{RF} \mathbf{f}_{u}^{BB} \mathbf{s}_{u} + \mathbf{n}_{u}
\end{equation}
 
where $\mathbf{h}_{n}$ denotes the channel between the transmitter and the \textit{n}th user. Since the mmW propagation environment has a limited number of scatterers, in this paper we follow the geometric Saleh-Valenzuela model to represent the low rank mmW channel \cite{ref_13}:
\begin{equation}\label{eq:chan}
    \mathbf{h}_{n} = \sqrt{\left(\dfrac{N_{T} N_{R}}{\epsilon L_{n}}\right)} \sum_{l=0}^{L_{n}}\eta_{n,l}\textbf{a}_{n,R}(\phi_{n,R},\theta_{n,R})\textbf{a}^{H}_{T}(\phi_{n,T},\theta_{n,T})
\end{equation}

$L_{n}$ represents the number of effective channel paths linked to the limited number of scatterers for each user, $\epsilon$ is the path-loss, $\eta_{n,l}$ is the complex gain associated with the \textit{l}th path, $\textbf{a}_{n,R}$ and $\textbf{a}_{n,T}$ are the spatial signatures of the receiver and the transmitter, respectively, $\phi_{R}$ and $\theta_{R}$ represent the elevation and azimuth angle of arrival (AoA) and $\phi_{n,T}$ and $\theta_{n,T}$ represent the elevation and azimuth angle of departure (AoD) of the \textit{l}th path at the receiver and at the transmitter, respectively. For uniform planar arrays (UPAs), the spatial signatures are estimated as given in \cite{ref_01}. 

The aim is to select $N_{r}$ users from a total of $N_{R}$ single antenna users so that the sum rate is maximized. The received signal to interference plus noise ratio (SINR) for the \textit{k}th user after precoding is given as
\begin{equation}
	\gamma_{k} = \frac{\left| \mathbf{b}_{k}\mathbf{F}_{RF} \mathbf{f}_{k}^{BB} \right|^2} {\sum_{\acute{k}\neq k} \mathbf{b}_{k}\mathbf{F}_{RF} \mathbf{f}_{\acute{k}}^{BB} + \sigma^2} 
\end{equation}

where $\sigma$ is the noise power and $\mathbf{b}_{k}$ is the channel between the transmitter and the \textit{k}th selected user. Furthermore, let $b_{k}$ represent the \textit{k}th row of $\mathbf{B}$ $\in C^{N_{r} \times N_{T}}$ which consists of rows of channel matrix $\mathbf{H}$ that correspond to the channel of selected users obtained using our proposed CNN architecture. The sum rate of the system after user selection is expressed as
\begin{equation}
	R = \log \bigg(\mathbf{I}_{N_{r}} + \sum_{k=1}^{N_{r}} \gamma_{k}\bigg)
\end{equation}
\begin{figure*}[htbp]
\centering
\includegraphics[height=1.6in,width=6.0in]{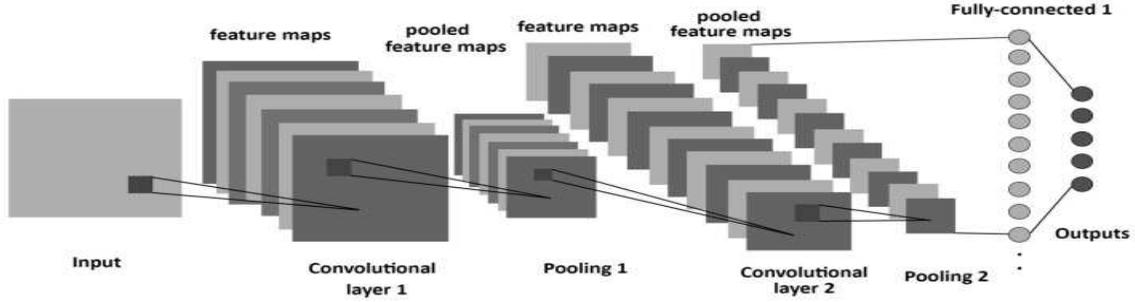}
\caption{Deep CNN Architecture}
\label{fig:DL}
\end{figure*}

\section{CNN Based User Selection}
In this section, we first introduce the dataset and the label generation which will serve as the input to the CNN model for training. Then, we will describe the CNN architecture in detail.
\subsection{Data Normalization}
A channel matrix realization is a data sample. Since we have assumed single antenna users, each row of $\mathbf{H}$ represents the channel for the respective user. Since the CNNs do not accept complex entries as input, complex entries ($h_{ij}$) of channel matrix $\mathbf{H}$ are divided into real and imaginary parts and concatenated into a single matrix.
\subsection{Label Generation}
In order to select $N_{r}$ users from the $N_{R}$ available users, we have $W$ = $N_{R} \choose N_{r}$ combinations. Every combination is represented as a class pattern, so we have a total of $W$ class labels. We have a one to one correspondence between each set of a selected users and a class label. For example, the first pattern of combinations of selected users is given label 1, the second pattern of combinations of selected users is given label 2, and so on.

In order to generate a dataset for the training of our CNN, an exhaustive search is performed on each channel realization. The search algorithm examines all possible $W$ combinations generated and selects the combination of $N_{r}$ users that maximizes the sum rate. Hence, each channel realization is labeled with a class and given as input for the training of the CNN.
\subsection{Deep CNN Architecture} \label{cnn}
We have adopted a LeNet architecture for our CNN as shown in Fig \ref{fig:DL}. There are 2 convolutional layers, 2 pooling layers, 1 fully connected layer and 1 soft-max output layer. The input of the CNN is the normalized channel matrix. The first convolutional layer filters the normalized input channel matrix with 16 kernels of size $3 \times 3$. Then the first pooling layer normalizes and pools the input into a $3 \times 3 \times 16$ output response. The max-pooling kernels have a size of $2 \times 2$ and a stride of 2. The second convolutional layer filters the response with 32 kernels of size $3 \times 3$. The second pooling layer converts the input response to a $2 \times 2 \times 32$ output response. The fifth layer is a dense fully-connected layer with 1024 fully-connected kernels of size $1 \times 1$, which is a dense layer accelerating the convergence. Then a dropout layer, which randomly resets the output of each hidden neuron to zero, is added behind the fully-connected layer in order to avoid overfitting. The output of our CNN is the soft-max layer, which produces a class label. There is a one-to-one correspondence between a class label and a user selection. For the convolutional and fully-connected layers the rectified linear unit (ReLU) is used as the nonlinear activation function. The cross entropy is employed as the loss function with gradient descent optimizer.
\section{Simulations and Results}
\begin{figure}[tbph]
\centering
\includegraphics[height=1.6in,width=2.4in]{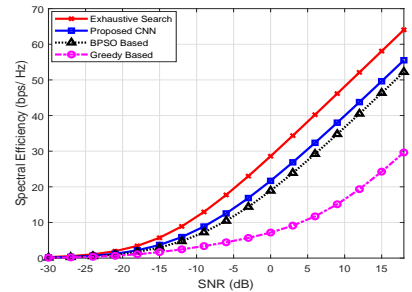}
\caption{Spectral Efficiency vs SNR for $N_{R}=10, N_{r}=6$}
\label{fig:res3}
\end{figure}

\begin{figure}[tbph]
\centering
\includegraphics[height=1.6in,width=2.4in]{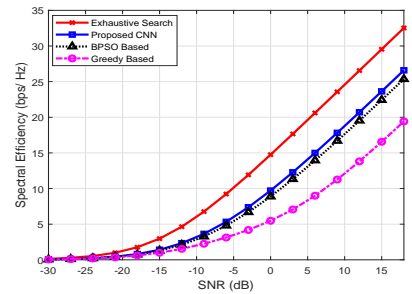}
\caption{Spectral Efficiency vs SNR for $N_{R}=10, N_{r}=3$}
\label{fig:res4}
\end{figure}

Python 3.6 with tensor flow is used for CNN training and prediction, whereas the comparison with model driven algorithms is performed in MATLAB. The dataset for the simulations is generated for 100,000 independent channel realizations. The number of transmit antennas $N_{T}$ is set to 144. An exhaustive search method is used to generate the class label for each channel realization. The number of epochs for the training of CNN is kept as 200. Batch size is kept as 100. Fig \ref{fig:res3} shows the result obtained by selecting 6 out of 10 active users and Fig \ref{fig:res4} shows the result obtained by selecting 3 out of 10 active users.

The results verify the performance of our CNN based algorithm over conventional iterative algorithms. It can be seen that the CNN based algorithm outperforms the sum rate achieved by the iterative binary particle swarm optimization (BPSO) and greedy algorithms. The greedy algorithm has the worst performance due the fact that it cannot select the “best” set of users from the available users. Although our proposed CNN approach only marginally outperforms the BPSO algorithm in terms of sum rate, it achieves a large gain in terms of reduction in computational complexity, as discussed in Section \ref{ca}. Even though the proposed CNN algorithm has a performance gap with respect to the exhaustive search algorithm, the latter is not suitable due to high computational complexity.
\subsubsection{Performance Analysis with Imperfect CSI}
\begin{figure}[tbph]
\centering
\includegraphics[height=1.6in,width=2.4in]{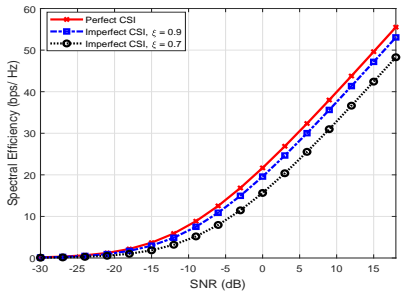}
\caption{Impact of imperfect CSI on the proposed system}
\label{fig:resd}
\end{figure}

Next we evaluate the case of uncertainties affecting the channel model. The estimated channel matrix with imperfect CSI $\mathbf{\widehat{H}}$ can be modeled as \cite{ref_02}
\begin{equation}
    \mathbf{\widehat{H}} = \xi\textbf{H} + \sqrt{1-\xi^2}\textbf{E} 
\end{equation}
where $\textbf{H}$ is the actual channel matrix, $\xi \in$ [0,1] is the CSI accuracy parameter and $\textbf{E}$ is an error matrix with entries following an i.i.d. $\mathbb{CN}$ (0,1). Fig. \ref{fig:resd} shows the effect of imperfect CSI on selecting 6 out of 10 active users with $\xi$ = 0.9 and 0.7. It is evident that the proposed system is not overly sensitive to CSI accuracy. It can also be seen that the achievable rate of the proposed system is quite close to that of the perfect CSI scenario even when $\xi = 0.7$.
\subsection{Complexity Analysis}\label{ca}
In this section, we perform a complexity analysis of the proposed CNN method and compare it with that of conventional model driven methods. We consider $N_{T}$ = 144 transmit antennas and  $N_{R}$ = 10 total available users. For our proposed CNN, the complexity in the offline training stage is normally not counted \cite{ref_14}, and only that in the online deployment stage is counted. With the parameters mentioned in Section \ref{cnn}, the total number of operations for CNN is around 6 million.

The exhaustive search algorithm needs to compute $N_{R} \choose N_{r}$ complex determinants to evaluate the sum rate and each determinant requires O($N_{T}^{2}N_{r}^{2}$) complex operations to compute the hybrid beamforming matrices ($\mathbf{F}_{RF}$ and $\mathbf{F}_{BB}$) \cite{ref_14}. So for $N_{T}$=144, $N_{R}$=10 and $N_{r}$=6, the total number of operations is around 156 million. The BPSO based algorithm requires $N_{pop} \times N_{iter}$ determinants, where $N_{pop}$ is the population size and $N_{iter}$ is the total number of iterations for the BPSO algorithm. Both variables are set to 10 during the simulations. So, the total number of operations is around 74 million for the BPSO algorithm. The greedy algorithm needs to compute O($N_{T}^{2}N_{r}^{2}$) complex operations for a total number of $N_{R}$ users. So, the total number of operations is around 7.5 million for the greedy algorithm.

It can be seen that the proposed CNN has competitive computational complexity when compared with traditional model-based algorithms. In addition, the main operation of any CNN based algorithms only involves large-scale matrix multiplications and additions, which can be effectively accelerated by using graphics processing units. Conversely, most traditional model-based algorithms involve serial iterations where the optimization of the next iteration depends on the result of the previous iteration, which is not suitable for parallel computing.   
\section{Conclusions}
We have proposed a CNN based design approach for user selection involving large antenna arrays. It is shown that the proposed CNN based approach is more efficient in terms of sum rate than model driven methods and performs close to the optimal exhaustive search based user selection. In addition, the computational complexity of data driven methods is acceptable and calculations only involve matrix multiplications and additions in the online phase.


\begin{thebibliography}{10b0}
\bibitem{ref_01}
O. El Ayach, et al., "Spatially sparse precoding in millimeter wave MIMO systems," \textit{IEEE Trans. Wireless Commun.}, vol. 13, no. 3, pp. 1499-1513, Mar. 2014.
\bibitem{ref_02}
X. Gao, et al., "Energy-efficient hybrid analog and digital precoding for mmWave MIMO systems with large antenna arrays," \textit{IEEE J. Sel. Areas Commun.}, vol. 34, no. 4, pp. 998-1009, Apr. 2016
\bibitem{ref_03}
A. Alkhateeb, et al., "Limited Feedback Hybrid Precoding for Multi-User Millimeter Wave Systems," in \textit{IEEE Transactions on Wireless Communications}, vol. 14, no. 11, pp. 6481-6494, Nov. 2015
\bibitem{ref_04}
M. Naeem and D. C. Lee, "A Joint Antenna and User Selection Scheme for Multiuser MIMO System," \textit{Applied Soft Computing}, vol. 23, pp. 366-374, Oct. 2014.
\bibitem{ref_05}
G. Dimic, et al., "On downlink beamforming with greedy user selection: Performance analysis and a
simple new algorithm," \textit{IEEE Trans. Signal Process.,} vol. 53, pp. 3857-3868, Oct. 2005.
\bibitem{ref_06}
J. Cui, et al., "User Selection and Power Allocation for mmWave-NOMA Networks," 2017 IEEE Global Communications Conference, Singapore, pp. 1-6.
\bibitem{ref_07}
Maimaiti, et al., "A low-complexity algorithm for the joint antenna selection and user scheduling in multi-cell multi-user downlink massive MIMO systems", \textit{J Wireless Com Network}, 2019
\bibitem{ref_08}
J. Choi, N. Lee, S. Hong and G. Caire, "Joint User Selection, Power Allocation, and Precoding Design With Imperfect CSIT for Multi-Cell MU-MIMO Downlink Systems," in \textit{IEEE Transactions on Wireless Communications}, vol. 19, no. 1, pp. 162-176, Jan. 2020.
\bibitem{ref_09}
H. Ye, G. Y. Li, and B. Juang, "Power of deep learning for channel estimation and signal detection in OFDM systems," \textit{IEEE Wireless Commun. Lett.}, vol. 7, no. 1, pp. 114-117, Feb. 2018.
\bibitem{ref_10}
H. He, C. Wen, S. Jin and G. Y. Li, "Deep learning-based channel estimation for beamspace mmWave massive MIMO systems," \textit{IEEE Wireless Commun. Lett.}, vol. 7, no. 5, pp. 852-855, Oct. 2018.
\bibitem{ref_11}
A. M. Elbir and K. V. Mishra, "Joint Antenna Selection and Hybrid Beamformer Design Using Unquantized and Quantized Deep Learning Networks," in \textit{IEEE Transactions on Wireless Communications}, vol. 19, no. 3, pp. 1677-1688, Mar. 2020
\bibitem{ref_12}
J. Joung, "Machine Learning-Based Antenna Selection in Wireless Communications," in \textit{IEEE Communications Letters}, vol. 20, no. 11, pp. 2241-2244, Nov. 2016.
\bibitem{ref_13}
A. Alkhateeb, et al., "Channel estimation and hybrid precoding for millimeter wave cellular systems", \textit{IEEE Journal of Selected Topics in Signal Processing}, vol. 8, no. 5, pp. 831-846, Oct. 2014.
\bibitem{ref_14}
X. Gao, et al., “ComNet: combination of deep learning and expert knowledge in OFDM receivers,” \textit{IEEE Commun. Lett}., vol. 22, no. 12, pp. 2627-2630, Dec. 2018.
\bibitem{ref_15}
Huang, et al., "DFT codebook-based hybrid precoding for multiuser mmWave massive MIMO systems". \textit{EURASIP J. Adv. Signal Process}. 2020.

\end{thebibliography}
\end{document}